\documentclass[conference]{IEEEtran}

\usepackage{color}
\usepackage[english]{babel}
\usepackage{booktabs}
\usepackage{paralist}
\usepackage{graphicx}
\usepackage{multirow}
\usepackage{soul}
\usepackage{amsmath}
\usepackage{cleveref}
\usepackage{url}
\usepackage{float}
\usepackage{cite}
\usepackage{enumerate}
\usepackage{enumitem}
\usepackage{fixltx2e}
\usepackage{algorithm}
\usepackage[noend]{algpseudocode}
\usepackage{relsize}
\usepackage{subfig}
\usepackage{setspace}
\usepackage{balance}
\usepackage{epstopdf}
\usepackage{balance}

\usepackage{booktabs}
\usepackage{amstext} 
\usepackage{array}   
\newcolumntype{L}{>{$}l<{$}} 

\usepackage{cleveref}
\crefformat{section}{\S#2#1#3}
\crefformat{subsection}{\S#2#1#3}
\crefformat{subsubsection}{\S#2#1#3}
\crefrangeformat{section}{\S\S#3#1#4 to~#5#2#6}
\crefmultiformat{section}{\S\S#2#1#3}{ and~#2#1#3}{, #2#1#3}{ and~#2#1#3}

\usepackage[breakable, skins]{tcolorbox}
\tcbset{arc=0mm,size=fbox,attach title to upper={\ ---\ },coltitle=black}

\newcommand{\RY}[1]{}

\def\BibTeX{{\rm B\kern-.05em{\sc i\kern-.025em b}\kern-.08em
    T\kern-.1667em\lower.7ex\hbox{E}\kern-.125emX}}

\begin{document}

\title{Coexistence of Age Sensitive Traffic and High Throughput Flows: Does Prioritization Help?}

\author{\IEEEauthorblockN{Tanya Shreedhar}
\IEEEauthorblockA{\textit{Wireless Systems Lab, IIIT-Delhi} \\
tanyas@iiitd.ac.in}
\and
\IEEEauthorblockN{Sanjit K. Kaul}
\IEEEauthorblockA{\textit{Wireless Systems Lab, IIIT-Delhi} \\
skkaul@iiitd.ac.in}
\and
\IEEEauthorblockN{Roy D. Yates}
\IEEEauthorblockA{\textit{WINLAB, Rutgers University} \\
ryates@winlab.rutgers.edu}
}

%
\maketitle
%
\IEEEpubidadjcol




\begin{abstract}
We study the coexistence of high throughput traffic flows with status update flows that require timely delivery of updates. A mix of these flows share an end-to-end path that includes a WiFi access network followed by paths over the Internet to a server in the cloud. Using real-world experiments, we show that commonly used methods of prioritization (DSCP at the IP layer and EDCA at the 802.11 MAC layer) in networks are highly effective in isolating status update flows from the impact of high throughput flows in the absence of WiFi access contention, say when all flows originate from the same WiFi client. Prioritization, however, isn't as effective in the presence of contention that results from the throughput and status update flows sharing WiFi. This results in prioritized status update flows suffering from a time-average age of information at the destination server that is about the same as when all flows have the same priority.  
\end{abstract}  


 \section{Introduction}
\label{sec:introduction}
IoT devices often communicate their updates, which require timely delivery to a server in the cloud, over an end-to-end path that includes a shared wireless access followed by a multihop path over the Internet to the server. The update traffic often shares the path with traffic that would like to achieve high throughput.
Update packets that require timeliness will suffer large delays if queued together with high throughput flows. They may also suffer significant delays in obtaining transmission opportunities over a shared multiaccess when competing for the same with high throughput flows. In practice, the networking stack allows priorities to be associated with data flows using, for example, the mechanism of Differentiated Services Code Point (DSCP). In principle, this can help alleviate the adverse consequences of update packets sharing the network with throughput flows.

In this work, we empirically shed light on the benefits of prioritizing update packets sent over a shared WiFi access to a server in the cloud. We use the time average age of information (AoI)~\cite{KaulYatesGruteser-Infocom2012} to quantify timeliness. Transmission Control Protocol (TCP) flows are used to emulate high throughput traffic. For end-to-end flows carrying update packets, we regulate the end-to-end rate of updates using the Age Control Protocol (ACP)~\cite{shreedhar2018acp, tanya-kaul-yates-wowmom2019,tanya-kaul-yates-arxiv,Shreedhar-AoIWorkshop2020-ACP+}, which has been shown to provide good timeliness performance over paths of interest in this work.

Work on optimizing metrics of the age of information has considered packet management techniques, including priorities and preemption when multiple sources share a service facility~\cite{Kaul-Yates-isit2018priority,Maatouk-Assaad-ISIT2019,Xu-Gautam-PriorityArXiv2019,Moltafet-2021-TCOM-PktMgmt,Huang-Modiano-isit2015,Najm-Telatar-INFOCOMWKS2018,Najm-Nasser-IT2020}. Such work often uses queueing models to capture sharing of the network resources. However, contention has not been modeled when sources share a multiaccess channel. Also, these works assume that all traffic sharing the facility requires timely delivery. Last but not least, it is often assumed that packet management may discard a source packet. 

Given the shared WiFi access and Enhanced Distributed Contention Access (EDCA), we have different queues for the ACP and TCP flows in our work as we assign a higher priority to update packets (ACP flows). The queues, however, are FCFS and don't allow preemption. Our specific contributions are:
\begin{enumerate}[leftmargin=*]
    \item We provide an empirical evaluation of the impact of coexisting ACP and TCP flows on the time-average age of information of the ACP flows and the throughputs of the TCP flows. Both flows share a WiFi network and have a server in the cloud as their destination.  
    \item Using different experimental configurations (a) with and without prioritization, (b) with and without shared access, and (c) in the absence of TCP flows, we show that while giving ACP flows higher priority in the absence of contention over the WiFi access (all flows originate from the same WiFi client) effectively isolates the ACP flows from the TCP flows, the contention that is caused when WiFi access is introduced and all flows originate from different WiFi clients results in barely any gains from prioritization.  
    \item We show from our experiments that as the number of ACP flows become large enough, TCP and ACP flows (prioritized) sharing the same WiFi access is worse both in terms of the throughputs of the TCP flows and the timeliness achieved by the ACP flows. 
\end{enumerate}
 \section{Related Work}
\label{sec:related}

Several works~\cite{Kaul-Yates-isit2018priority,Maatouk-Assaad-ISIT2019,Xu-Gautam-PriorityArXiv2019,Moltafet-2021-TCOM-PktMgmt} analyze the average age of updates in the presence of priority traffic.
In~\cite{Kaul-Yates-isit2018priority}, the authors analyze the average age of updates when the sources are assigned different priorities for two service facilities. One which allows source agnostic preemption in service by a new arrival of equal or higher priority and the other in which there is a waiting room of size $1$ and a new arrival can preempt an update in waiting but not in service. In~\cite{Maatouk-Assaad-ISIT2019}, the authors expand the waiting room to allow each source to have up to one waiting update while the server is busy. This also allows an update in service that is preempted by a higher priority source to be saved in the waiting room to resume service later. In~\cite{Xu-Gautam-PriorityArXiv2019}, the authors analyze peak age when sources have priorities and queues are of infinite size for Poisson arrivals and general service times. In~\cite{Moltafet-2021-TCOM-PktMgmt}, the authors propose and analyze three source aware packet management policies considering a memoryless service facility of a single queue and server. The facility sees arrivals from two independent Poisson sources. The policies make different choices regarding the size of the waiting room and whether preemption is allowed in the service. In~\cite{Najm-Telatar-INFOCOMWKS2018}, the update currently in service is preempted instead of discarding a new arrival. In~\cite{Najm-Nasser-IT2020} arrivals consist of a mix of ordinary and priority updates. The latter can preempt any update in service. In case the preempted update is ordinary, it is not discarded and is queued for resuming service later.

Systems research that analyzes AoI in real-world settings is relatively limited~\cite{Sonmez-BaghaeeSeaCom2018,shreedhar2018acp,tanya-kaul-yates-wowmom2019,tanya-kaul-yates-arxiv,Shreedhar-AoIWorkshop2020-ACP+,kadota2020wifresh}. \cite{Sonmez-BaghaeeSeaCom2018} brings forth the need for an AoI optimizer that can adapt to changing network topologies and delays. The Age Control Protocol (ACP)~\cite{shreedhar2018acp,tanya-kaul-yates-wowmom2019,tanya-kaul-yates-arxiv}, is a transport-layer solution that works in an application-independent and network-transparent manner and attempts to minimize the age of information of a source at a monitor connected via an end-to-end path over the Internet. In~\cite{Shreedhar-AoIWorkshop2020-ACP+}, the authors propose a modification to ACP and also compare it with other state-of-the-art TCP congestion control algorithms used in the Internet. In~\cite{kadota2020wifresh}, WiFresh, a MAC and application-layer solution to ageing of updates over a wireless network is proposed. There are also works on various other aspects of the metrics of AoI, including optimizing age over multiaccess channels and can be found in~\cite{kosta2017-FTN} and~\cite{yates2020age-survey}. 
\section{Prioritization in Networks}
\label{sec:priority}

\begin{figure}[!t]             
\begin{center}
\includegraphics[width=0.45\textwidth]{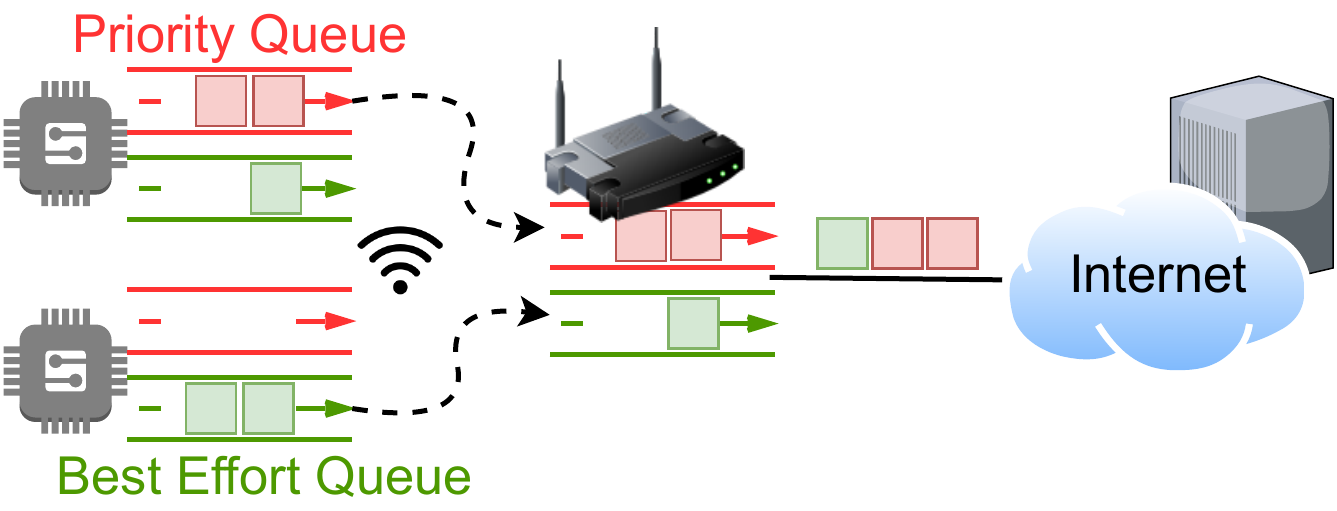}
\caption{\small Illustration of priority queueing over a shared WiFi access. Nodes and AP maintain separate queues for different service classes.}
\label{fig:priority_queue}
\end{center}
\end{figure}

Several mechanisms exist in modern networked systems that commonly address network bottlenecks by allowing priority packets to pass first~\cite{rfc4594,dscp}. 
The majority of such mechanisms operate by categorizing network traffic into distinct ``service classes" -- each one assigned a separate queue. Based on the QoS demands of each class, these mechanisms manage the rate of each class queue such that the services can access a bottleneck link depending on their priority.
For example, a router at the bottleneck link may handle voice-over-IP (VoIP) application traffic using a high-priority, high-rate queue, while packets of video streaming applications over the same link might be forwarded at a significantly lower rate.

There are several ways in which network operators can classify network flows into different service classes in their managed routers. For example, operators may use the destination IP address and port to identify an application type  (e.g., data egress from Netflix servers) or prioritize based on the transport protocol used (RTP may have a different priority than UDP/TCP traffic)~\cite{cisco}.
The most common traffic classification method uses Differentiated Services Code Point (DSCP) markings. Application providers can assign their packets with a unique code in the IP layer. Each code maps to a unique traffic class type that can be treated with a different priority. The current standard dictates network management control traffic to be assigned the highest priority, followed by interactive applications, low-loss low-latency data transfers, and finally, best-effort data transfer applications~\cite{rfc4594,dscp}.
As the DSCP value is embedded in the IP header (layer $3$) of every packet, it is visible to every router on the Internet and thus allows for end-to-end flow prioritization.  

However, since multiaccess schemes like WiFi operate at layer $2$ (medium access control) of the networking stack, they remain oblivious to DSCP markings in the IP layer. Instead, the 802.11 standard employs its prioritization using Enhanced Distributed Channel Access (EDCA) or Hybrid Controlled Channel Access (HCCA)~\cite{QoSinIEEE}.
Similar to DSCP, the 802.11 prioritization assigns eight separate queues at the MAC layer in which data packets are segregated based on their priority level (defined as \texttt{User Priority}). Each priority level is assigned to one of the four access categories (analogous to DSCP traffic classes), i.e. background (\texttt{AC\_BK}), best-effort (\texttt{AC\_BE}), video (\texttt{AC\_VI}) and voice (\texttt{AC\_VO}) (arranged in increasing priority order).
Each access category uses different CSMA/CA minimum and maximum contention window sizes and also inter-frame spacing (IFS). This enables packets belonging to a higher priority access category faster access to the shared channel and less contention from lower priority packets awaiting access.

Recent efforts have mapped DSCP markings to 802.11 EDCA priority and access categories~\cite{rfc8325}.
It is now possible for application providers to assign their traffic higher priority in both wired and wireless networks by setting DSCP in the IP header.
Table~\ref{tbl:diffserv} summarizes different traffic classes and their priority mappings between DSCP (Diffserv) and 802.11.

\section{Experimental Setup and Methodology}
\label{sec:methodology}

\begin{figure}[!t]             
\begin{center}
\includegraphics[width=0.49\textwidth]{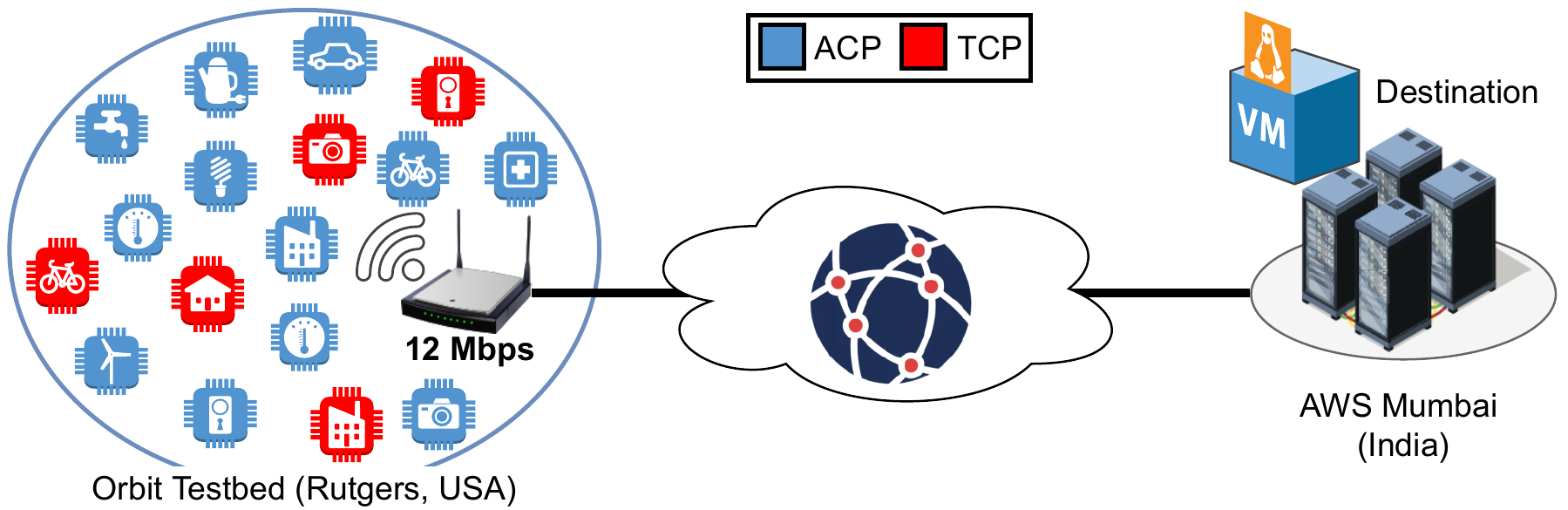}
\caption{An illustration of the network topology. Clients containing a mix of TCP (red) and ACP (blue) are connected to a WiFi AP located in the Orbit Testbed's WiFi grid in USA. The server is located in AWS Mumbai, India.}
\label{fig:orbit_setup}
\end{center}
\end{figure}

\Cref{fig:orbit_setup} illustrates our real-world experimental setup.
For our experiments, we use the ORBIT testbed~\cite{orbit}, which is an open wireless network emulator grid located in Rutgers University, USA. 
ORBIT houses multiple programmable radio nodes deployed in a controlled grid-like environment with adjacent WiFi nodes along rows and columns at a distance of $1$ m from each other. 
Each ORBIT node runs Ubuntu $18.04$ LTS over Linux kernel v$5.0$.
By default, ORBIT nodes use the $1$ Gbps ethernet NIC to connect to the Internet.
We set up one of the ORBIT nodes as an $802.11$n access point configured to operate at $5$ GHz on a fixed channel using \texttt{hostapd} and the Atheros (\texttt{ath9k}) Linux WiFi driver~\cite{hostapd}.
To focus on the interplay between priorities and contention, we disable the automated WiFi physical layer (PHY) rate control in \texttt{ath9k} drivers and instead use a \textit{fixed} WiFi PHY rate for the length of an experiment. While most of our experiments use a PHY rate of $12$ Mbps, we also use $6$ Mbps for some experiments. We provide experiment-specific PHY rates in \Cref{sec:evaluation}.
We select up to 80 nodes as WiFi clients in the ORBIT testbed and associate them to the ORBIT node configured as the WiFi access point.
Our WiFi access point routes every packet received over WiFi to the public Internet over Ethernet. 
We also set up a node in the testbed as a \textit{sniffer} that captures all packets sent over the WiFi channel during our experiments. The sniffer data allows us to estimate MAC layer packet retry percentages suffered by the ACP and TCP flows over the WiFi access. It also helps confirm that EDCA priorities have been applied.

We use an ec2 AWS instance in Mumbai, India, as our destination server for all flows.
The baseline round-trip-time (RTT), calculated by sending one packet for every ACK between our WiFi clients and the server is $\approx$ $200$-$210$ ms. 
We evaluate three different flow configurations.
\begin{enumerate}
    \item \texttt{ACP-default}. Update packets are sent over an end-to-end path between a WiFi client (the ACP source) and the AWS server (ACP monitor) using the Age Control Protocol~\cite{Shreedhar-AoIWorkshop2020-ACP+,acpGitRepo}. Update packets sent by ACP are given the default priority and treated as best-effort traffic. 
    
    \item \texttt{ACP-priority}. It is same as \texttt{ACP-default} but here ACP packets are given the highest network priority by setting the DSCP value as CS7 (see \Cref{tbl:diffserv}).
    
    \item \texttt{TCP-iperf}. We use \texttt{iperf3} to generate TCP traffic from WiFi clients to the AWS server. We configure each TCP flow to use the \texttt{cubic} congestion control~\cite{ha2008cubic}. TCP flows are always treated as best effort in our work. 
\end{enumerate}

\begin{table}[t]
\centering
\small
\resizebox{0.9\columnwidth}{!}{%
\begin{tabular}{@{}cccc@{}}
\toprule
\begin{tabular}[c]{@{}c@{}}\textbf{IETF Diffserv}\\ \textbf{Service Class}\end{tabular}           & \textbf{DSCP}                                                       & \begin{tabular}[c]{@{}c@{}}\textbf{802.11 Access}\\ \textbf{Category}\end{tabular} & \begin{tabular}[c]{@{}c@{}}\textbf{User} \\ \textbf{Priority}\end{tabular} \\ \midrule
Network Control                                                                 & CS7, CS6                                                   & \begin{tabular}[c]{@{}c@{}}\texttt{AC\_VO}\\ (Voice)\end{tabular}         & 7                                                        \\
Signaling                                                                       & CS5                                                        & \begin{tabular}[c]{@{}c@{}}\texttt{AC\_VI}\\ (Video)\end{tabular}         & 5                                                        \\
\begin{tabular}[c]{@{}c@{}}Multimedia \\ Conferencing/\\ Streaming\end{tabular} & \begin{tabular}[c]{@{}c@{}}AF41-43,\\ AF31-33\end{tabular} & \begin{tabular}[c]{@{}c@{}}\texttt{AC\_VI}\\ (Video)\end{tabular}         & 4                                                        \\
\begin{tabular}[c]{@{}c@{}}High Throughput \\ Data\end{tabular}                 & AF11-13                                                    & \begin{tabular}[c]{@{}c@{}}\texttt{AC\_BE}\\ (Best Effort)\end{tabular}   & 3                                                        \\
Low-Priority Data                                                               & CS1                                                        & \begin{tabular}[c]{@{}c@{}}\texttt{AC\_BK}\\ (Background)\end{tabular}    & 1                                                        \\ \bottomrule
\end{tabular}
}
\caption{\small Diffserv QoS mapping in wired (DSCP) and WiFi access.
} \label{tbl:diffserv}
\end{table}

\begin{figure*}[tb]             
\begin{center}
\subfloat[\small Two ACP flows]{\includegraphics[width=.25\linewidth]{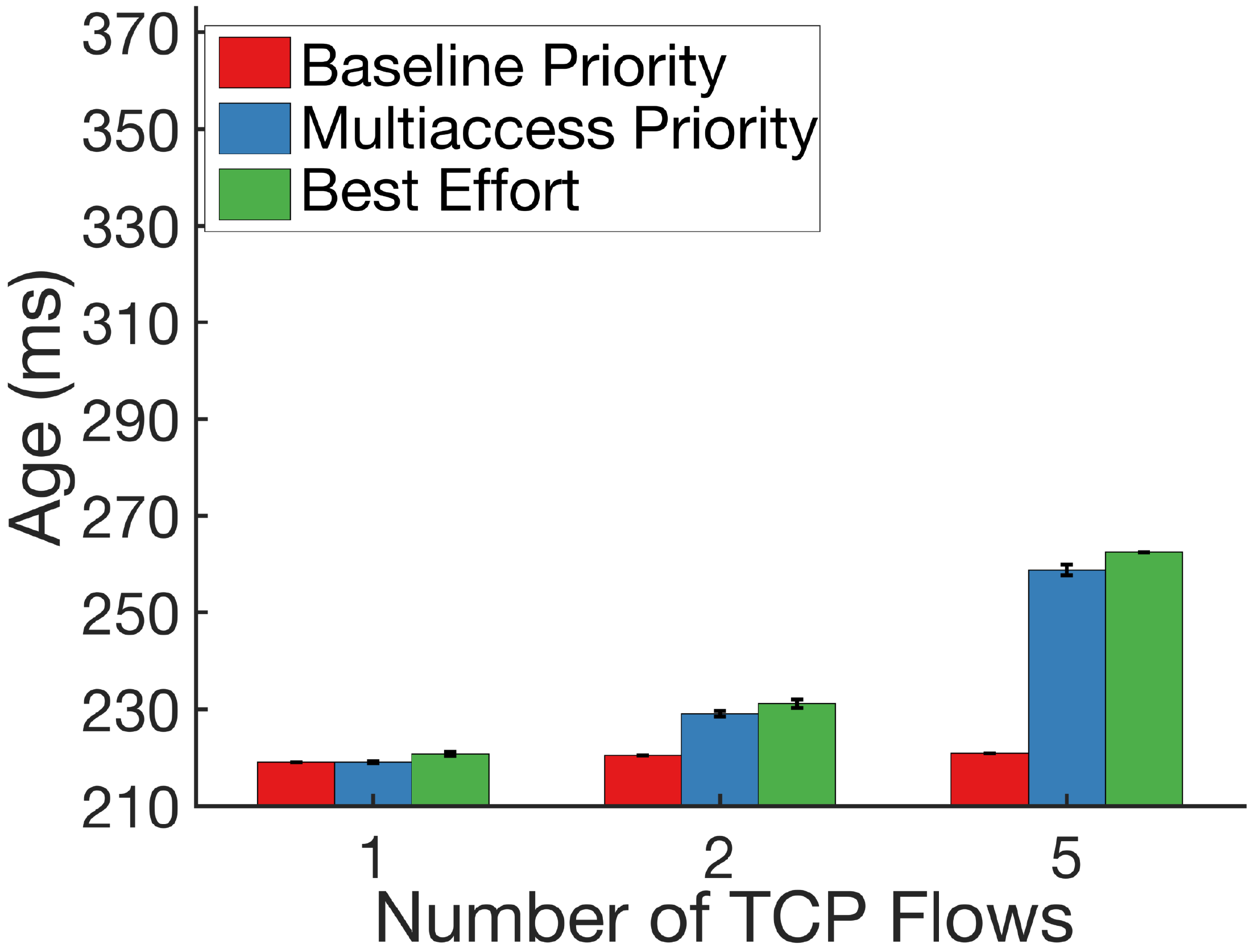}\label{fig:ACPAgeTwoFlows}}
\subfloat[\small Five ACP flows]{\includegraphics[width=.25\linewidth]{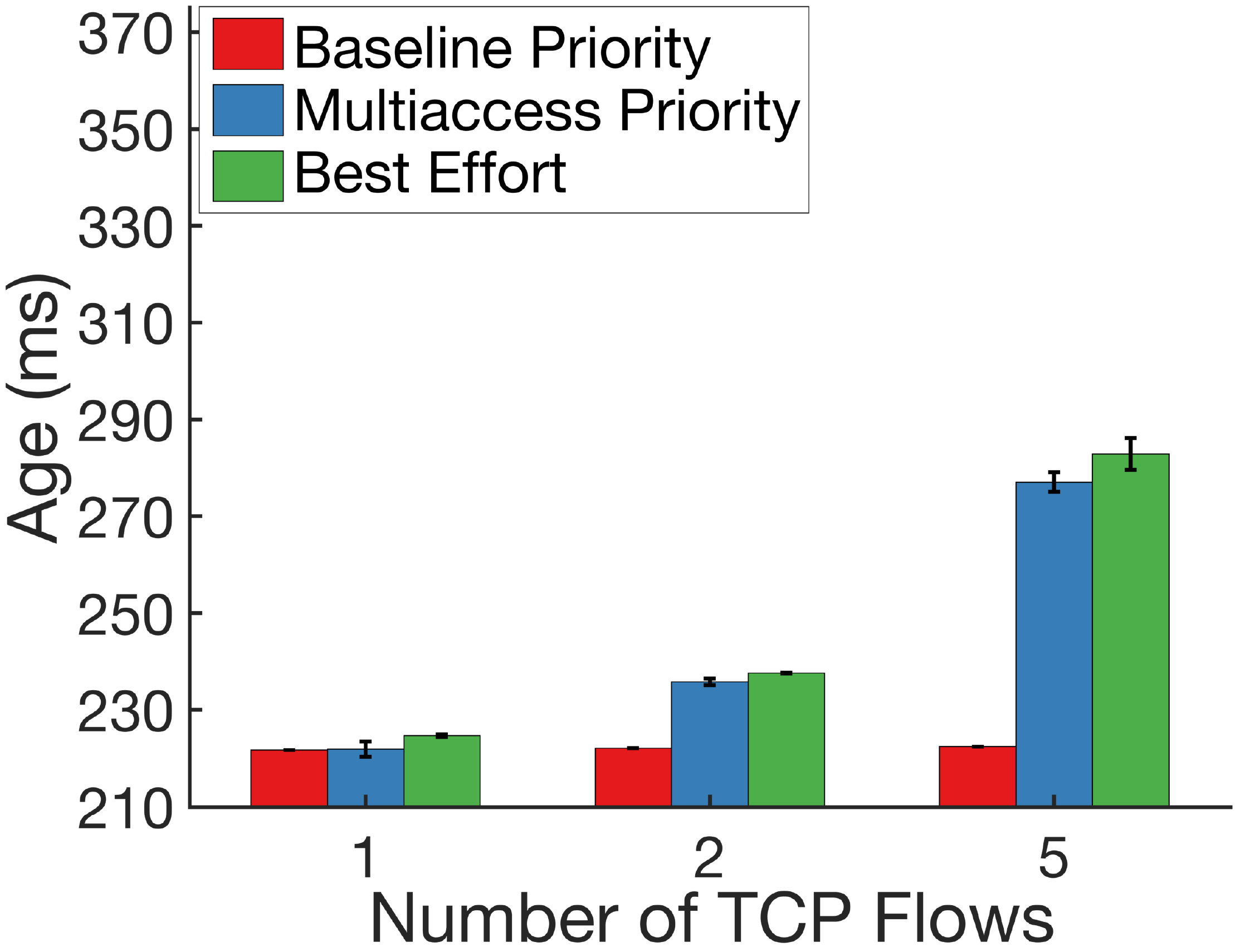}\label{fig:ACPAgeFiveFlows}}
\subfloat[\small Ten ACP flows]{\includegraphics[width=.25\linewidth]{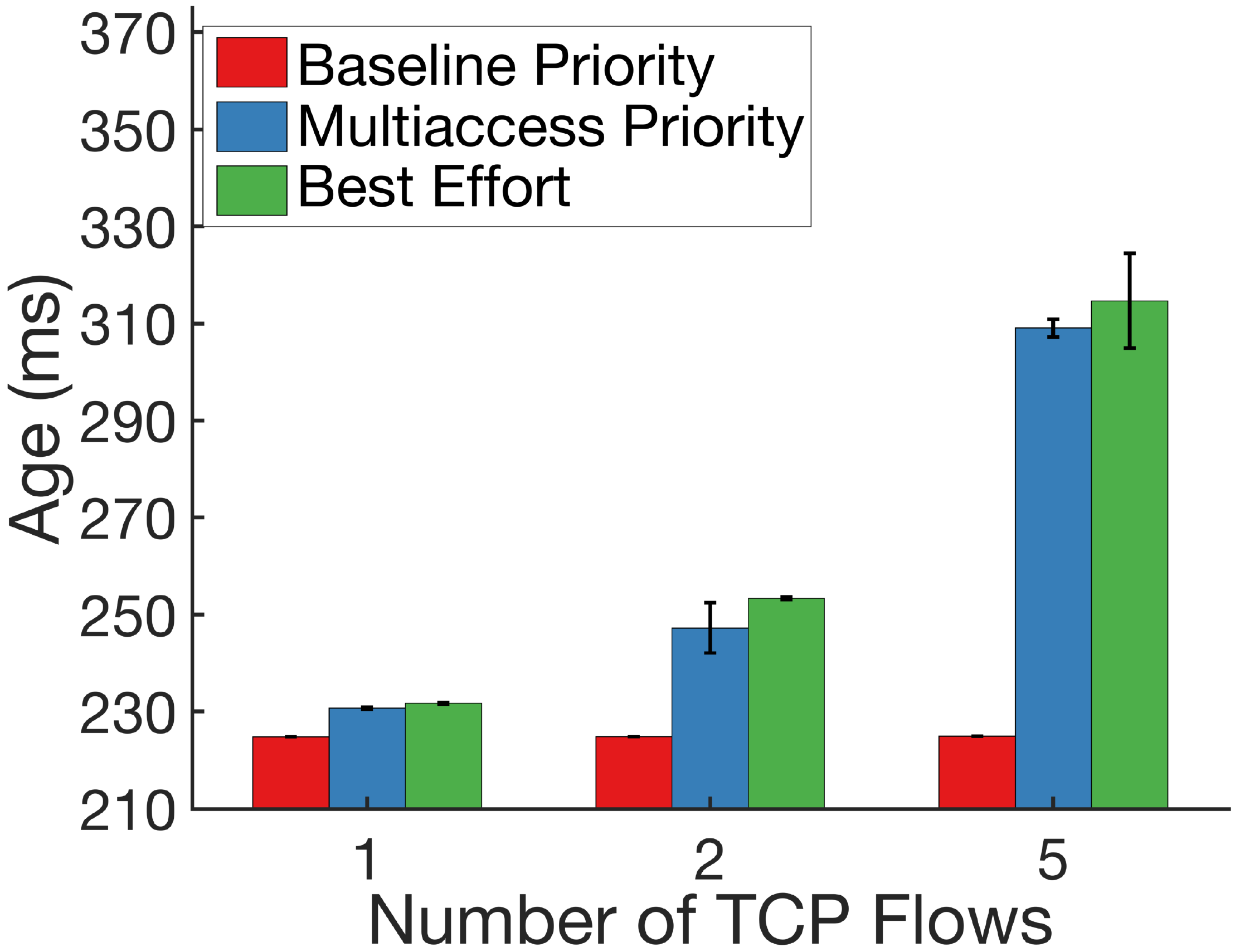}\label{fig:ACPAgeTenFlows}}
\subfloat[\small Twenty ACP flows]{\includegraphics[width=.25\linewidth]{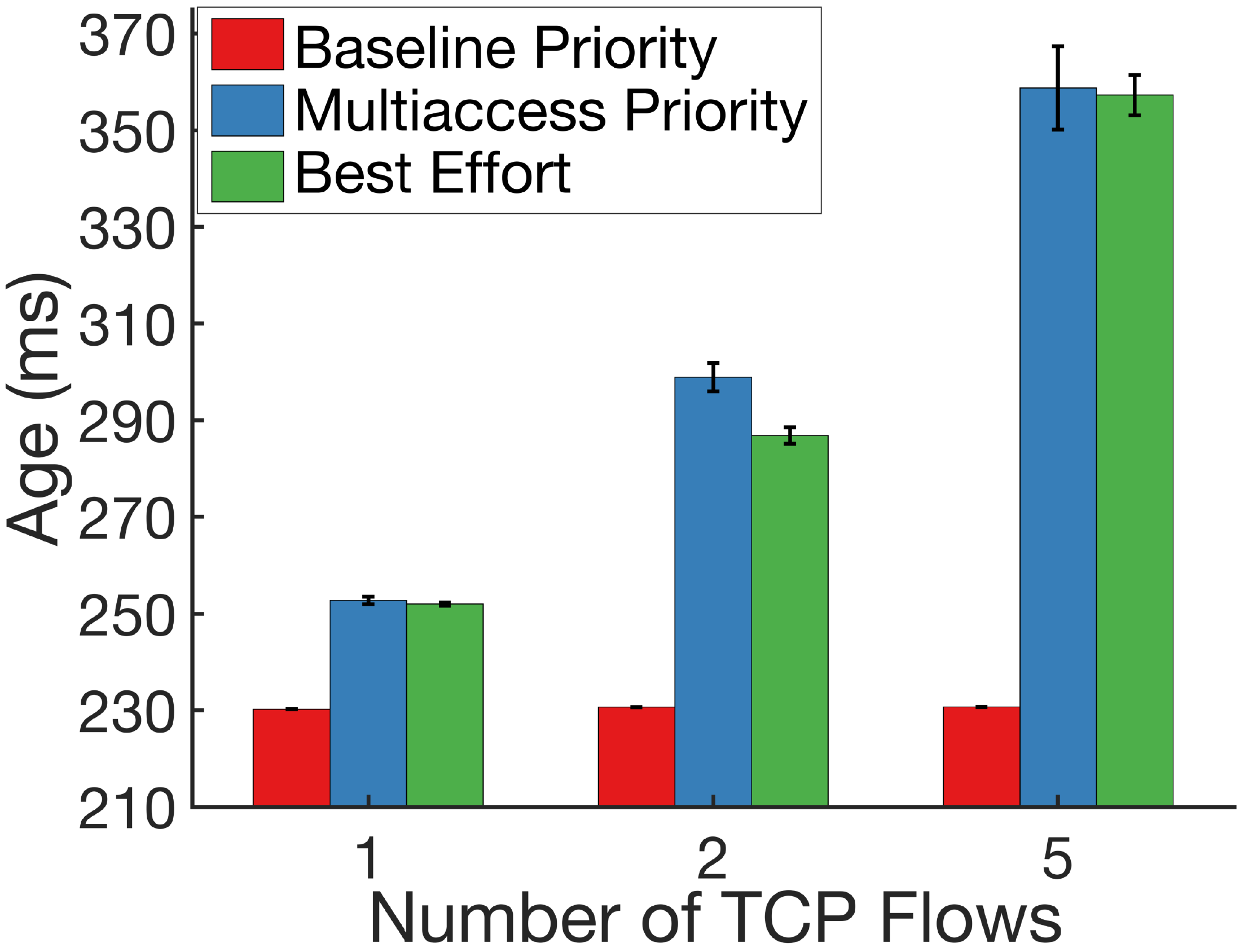}\label{fig:ACPAgeTwentyFlows}}
\caption{\small Mean time-average age achieved by $2$, $5$, $10$ and $20$ ACP flows for \textit{Baseline Priority}, \textit{Multiaccess Priority} and \textit{Best Effort} configurations for $1$, $2$, and $5$ coexisting TCP flows. 
}
\label{fig:ACPAge}
\end{center}
\end{figure*}

In addition to priority queueing at our configured WiFi access (available default in the 802.11 standard), we use \texttt{CAKE}~\cite{hoiland2018piece} at our AP node to support QoS priority at the Ethernet interface.
\texttt{CAKE} is a comprehensive network queue management utility that has been deployed as part of the OpenWRT framework and is available in all Linux kernels version $4.19$ and later~\cite{cake}.
\texttt{CAKE} supports Differentiated Services (DiffServ) prioritization scheme and maps \texttt{ACP-priority} flows to the highest priority queue (reserved for voice applications) ingressing the Ethernet interface.
On the other hand, \texttt{CAKE} treats flows belonging to \texttt{ACP-default} and \texttt{TCP-iperf} as the lowest priority best-effort traffic.
 
We use three different experiment configurations and simultaneously run ACP and TCP flows to evaluate the gains from prioritizing ACP flows.
Specifically, in \textit{Baseline Priority} (BP) we initiate one or more \texttt{ACP-priority} and \texttt{TCP-iperf} flows from a single WiFi client.
This setting eliminates any interference between the flows due to contention over the WiFi access. It focuses on the co-existence of ACP prioritized flows and TCP flows in what effectively is a setting with a single server and one FCFS queue for every priority class.
In \textit{Multiaccess Priority} (MP), each \texttt{ACP-priority} and \texttt{TCP-iperf} flow runs on a separate WiFi client. The flows therefore compete for the shared WiFi multiaccess, resulting in contention.
Lastly, in \textit{Best Effort} (BE), as in MP, each flow begins in a different WiFi client. We have \texttt{ACP-default} flows where no priority is assigned along with \texttt{TCP-iperf} flows. 
All flows are thus treated as best effort.

\smallskip
\noindent \textbf{Evaluation Metrics.}
We define our evaluation metrics. The performance of an ACP flow (\texttt{ACP-default} or \texttt{ACP-priority}) is evaluated in terms of the estimate of \emph{time-average age}~\cite{KaulYatesGruteser-Infocom2012} at the source.
Note that since the source of the flow (a WiFi client) is not time-synchronized with the AWS server, age can't be calculated at the server. We bank on ACP \texttt{ACK} packets sent by the server back to the source in response to every update packet sent by the source to estimate the age. The round-trip-time (RTT) corresponding to an \texttt{ACK-ed} source packet is assumed to be the packet's system time.
Age is assumed to reset to this time when the source receives an \texttt{ACK}. \textit{Out-of-sequence} older \texttt{ACKs} are discarded, which is in line with the \textit{freshness} requirement.
Using RTTs as an estimate of system time can lead to over-estimation of age. However, since we consider a linear age function, the bias in estimation does not affect the optimal operating point. Works~\cite{tanya-kaul-yates-wowmom2019,Shreedhar-AoIWorkshop2020-ACP+} contains the design principles and details about ACP. In \cref{sec:evaluation} we present the \emph{mean time-average age}, which is the mean calculated over all ACP flows.

We also discuss ACP throughput, which is the end-to-end rate (Mbps) of \texttt{ACK-ed} source packets and is calculated by the source based on the \texttt{ACK} packets it receives and the size of sent update packets. An update packet is of size $1024$ bytes in all our experiments. We will also present \emph{WiFi MAC packet retry percentages}. These simply capture the percentage of packets on air that were \textit{retries} for a source. 
The retry packets are marked with a retry flag which is captured by the sniffer.
Last but not least, TCP throughput is also a metric of interest. For all metrics, we present the mean calculated over $3$ repeats of an experiment, where each experiment is $1000$ seconds long. 
\section{Evaluation} \label{sec:evaluation}
We discuss our observations from experiments performed using the methodology described in~\cref{sec:methodology}. They help gain insight into whether prioritizing benefits ACP flows when they share a WiFi access with TCP flows. 

\smallskip
\noindent \emph{Are there any gains from prioritizing ACP flows over the shared WiFi multiaccess?}

Figure~\ref{fig:ACPAge} shows the mean time-average age achieved by ACP flows sharing the WiFi network with TCP flows. Figures~\ref{fig:ACPAgeTwoFlows},~\ref{fig:ACPAgeFiveFlows},~\ref{fig:ACPAgeTenFlows} and~\ref{fig:ACPAgeTwentyFlows} show the mean age for when the number of ACP flows are set to $2$, $5$, $10$, and $20$ respectively. For each selection of a number of ACP flows, the mean age is shown for when the number of TCP flows are $1$, $2$, and $5$. Further, for each selection of number of ACP and TCP flows, the mean is shown for the network configurations of \emph{Baseline Priority} (BP), \emph{Multiaccess Priority} (MP) and \emph{Best Effort} (BE).

Contention over the shared WiFi multiaccess increases in the configurations MP and BE when the number of ACP clients or TCP clients increases. Let's begin by considering the mean age achieved under \emph{Baseline Priority}. For a given number of ACP flows, for example, $5$ flows in Figure~\ref{fig:ACPAgeFiveFlows}, the mean age stays unchanged for different numbers of TCP flows. For $5$ ACP flows, this age is $\approx 222$ ms for $1$, $2$, and $5$ TCP flows. The age is $\approx 230.5$ ms when there are $20$ ACP nodes as in Figure~\ref{fig:ACPAgeTwentyFlows}. 
The age stays the same for a given number of ACP nodes because in BP all ACP and TCP flows originate in the same WiFi client. Because of their higher priority than TCP flows, ACP flows are unaffected by changes in the number of TCP flows originating in the WiFi client. On the other hand, an increase in the number of ACP flows does result in an increase in the mean age. This is because updates from a larger number of ACP flows share the same priority queue in the WiFi client. Mean age increases from $\approx 220$ ms for $1$ - $2$ ACP flows (Figure~\ref{fig:ACPAgeTwoFlows}) to $\approx 230$ ms for $20$ ACP flows (Figure~\ref{fig:ACPAgeTwentyFlows}).
\begin{figure*}[tb]             
\begin{center}
\subfloat[\small Two ACP flows]{\includegraphics[width=.21\linewidth]{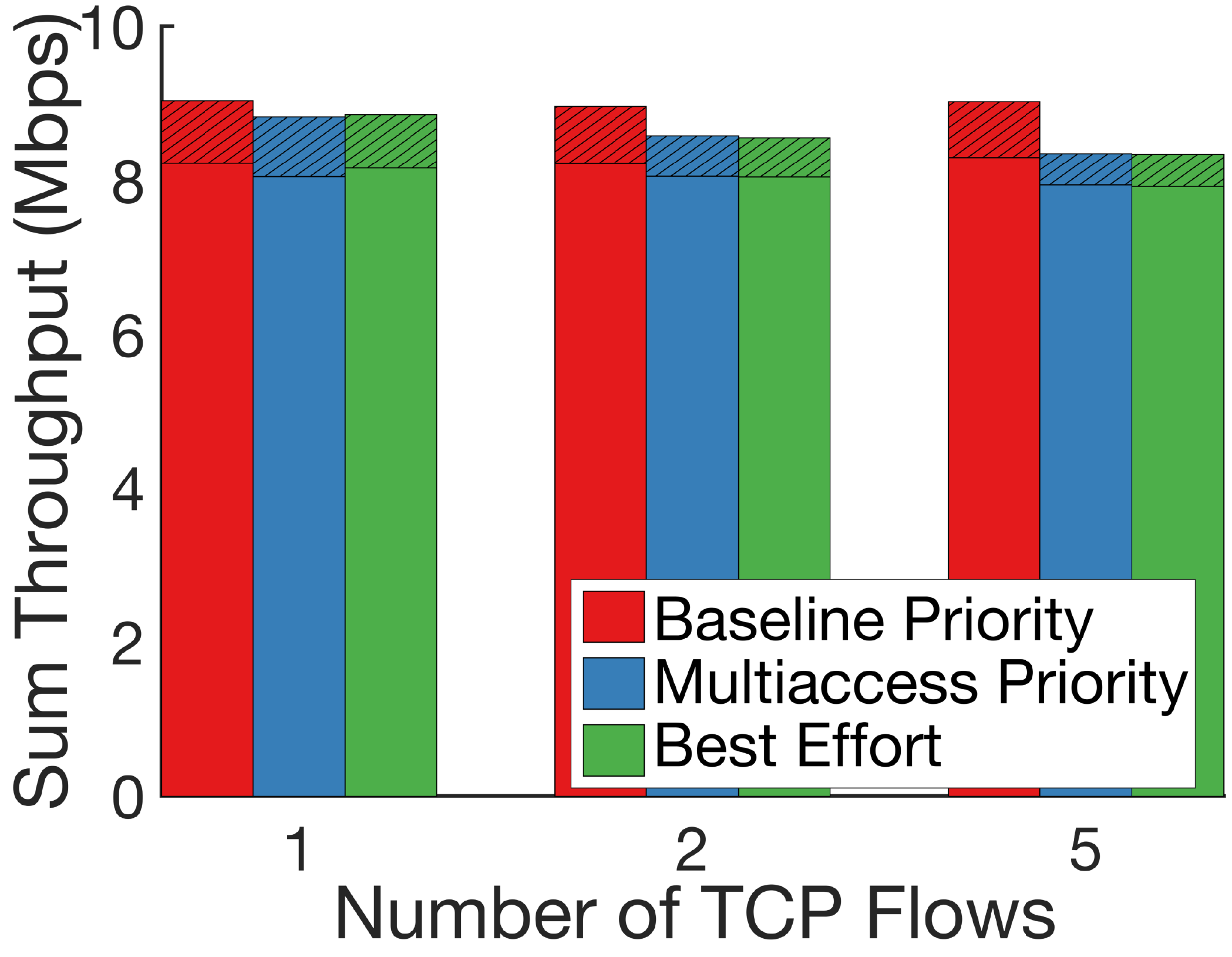}\label{fig:SumThrTwoFlows}}
\hspace{0.2in}
\subfloat[\small Five ACP flows]{\includegraphics[width=.21\linewidth]{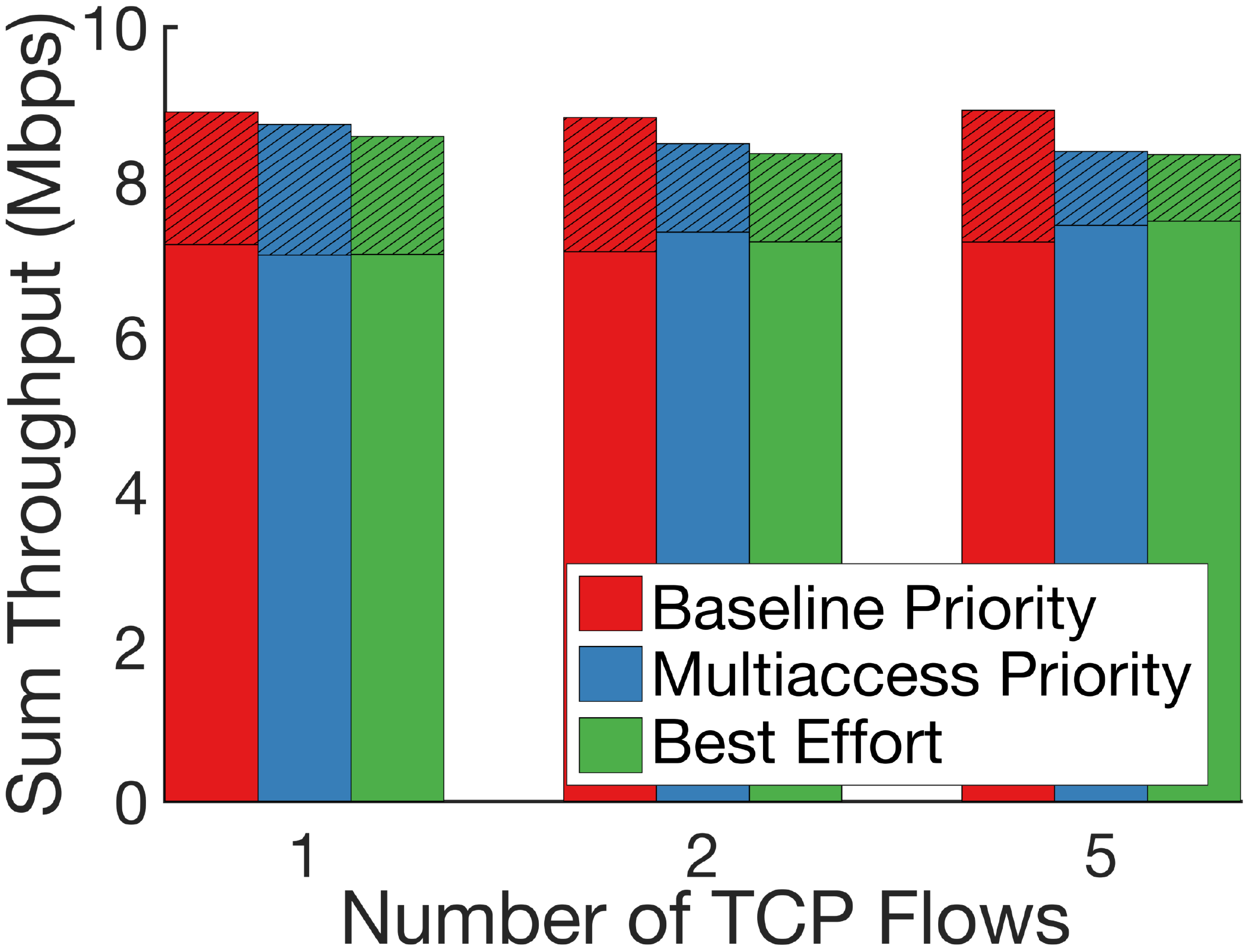}\label{fig:SumThrFiveFlows}}
\hspace{0.2in}
\subfloat[\small Ten ACP flows]{\includegraphics[width=.21\linewidth]{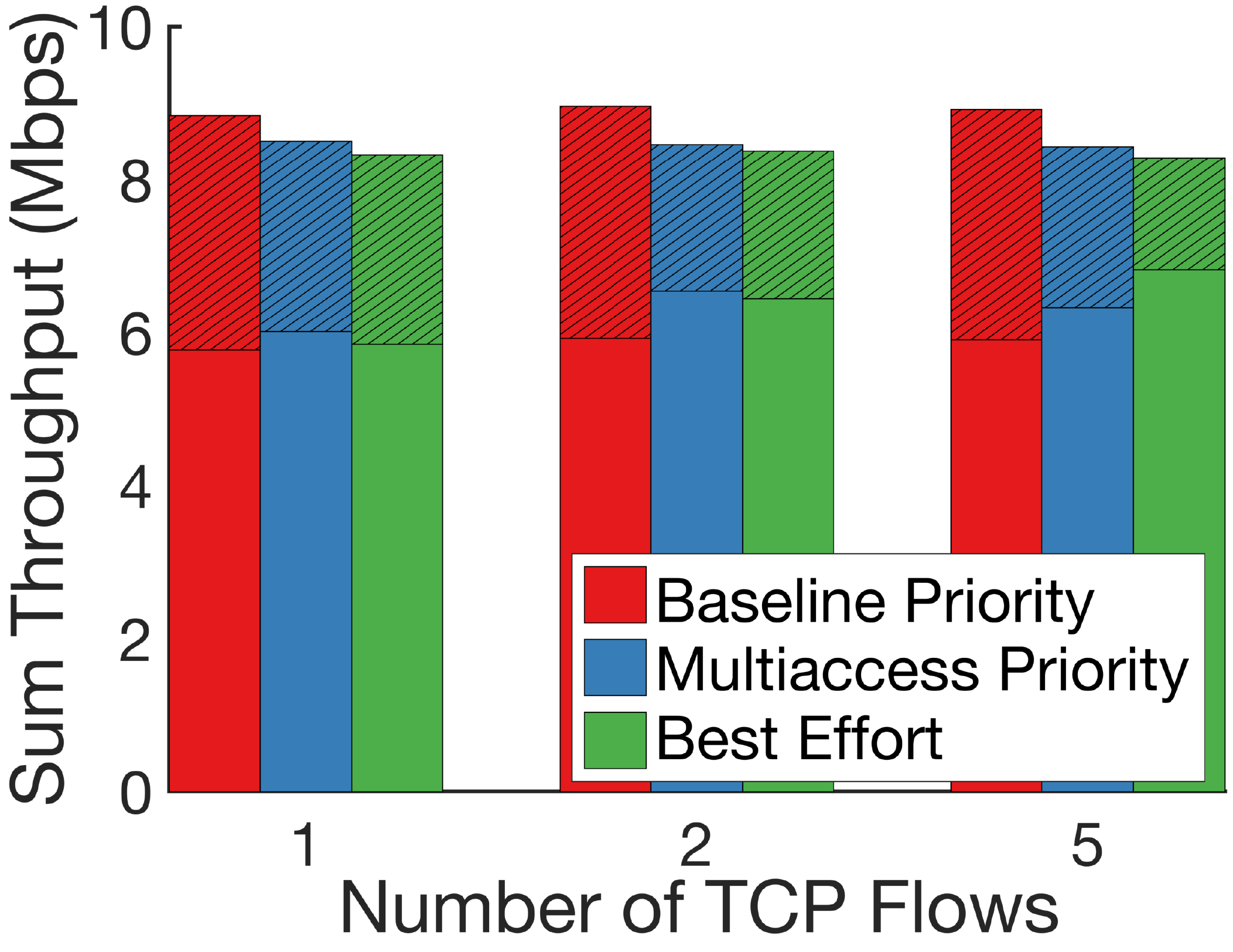}\label{fig:SumThrTenFlows}}
\hspace{0.2in}
\subfloat[\small Twenty ACP flows]{\includegraphics[width=.21\linewidth]{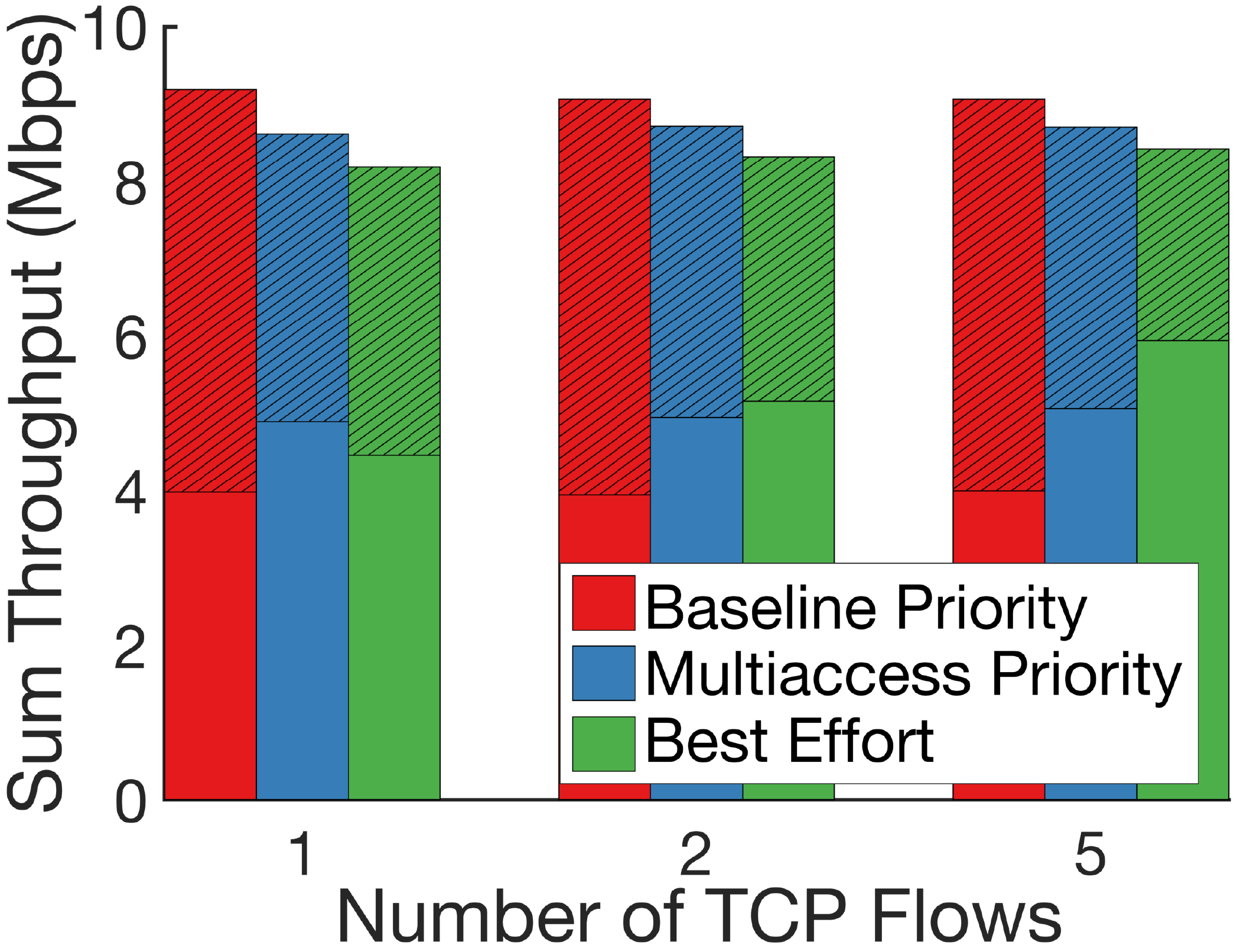}\label{fig:SumThrTwentyFlows}}
\caption{\small Sum throughput of ACP and TCP flows with their respective shares for $2$, $5$, $10$ and $20$ ACP flows. For each stacked bar, the diagonally striped top part corresponds to the sum of ACP flows and the bottom part shows the sum TCP throughput. For each number of ACP flows, the throughputs are shown for $1$, $2$, and $5$ coexisting TCP flows and for \emph{Baseline Priority}, \emph{Multiaccess Priority} and \emph{Best Effort}.}
\label{fig:SumThr}
\end{center}
\vspace*{-1em}
\end{figure*}
As seen in Figure~\ref{fig:ACPAge} 
for MP and BE configurations in which flows are distributed over different WiFi clients sharing the WiFi multiaccess, mean age increases significantly as the number of TCP flows increases for any selection of the number of ACP flows. 
Assigning a higher priority to ACP flows, as in MP, is ineffective in isolating them from the effects of TCP flows sharing the multiaccess. 
Also, the mean age when using \emph{Multiaccess Priority} is in general not much smaller than when treating ACP with the same priority as TCP when using \emph{Best Effort}. 

Further, we observe from Figure~\ref{fig:ACPAge} that for any selected number of TCP and ACP flows, contention over the multiaccess results in a higher mean age in comparison to BP. In fact, even for just $2$ ACP and $2$ TCP flows, we see that the mean age for the setting of MP is about $9$ ms more than that for \emph{Baseline Priority}. This gap increases rapidly with an increase in the number of ACP and TCP flows. For example, it jumps to $38$ ms for $2$ ACP and $5$ TCP flows, is $55$ ms for when we have $5$ ACP and $5$ TCP flows, and is $130$ ms for $20$ ACP and $5$ TCP flows.
To understand the reason behind significantly worse mean age when using MP, we begin by considering the throughput obtained by the TCP flows. Later we also look at the WiFi MAC layer retry percentages suffered by update packets of ACP flows and also their round-trip times (RTTs).

Figure~\ref{fig:SumThr} shows the sum throughput (sum of throughputs of all ACP and TCP flows) for the network configurations BP, MP, and BE, and different numbers of TCP and ACP flows. It can be observed that the sum throughput is about the same for all the configurations and all numbers of TCP and ACP flows. It stays in the very narrow range of $8.8$ to $9$ Mbps. Essentially, the TCP and ACP flows together achieve the available data payload rate of about $9$ Mbps, given the link rate of $12$ Mbps. The figure also shows the share of ACP flows and that of TCP flows in the sum throughput. As can be seen, the fraction of sum throughput that corresponds to ACP flows increases with the number of ACP flows. As expected, the sum throughput of ACP flows for \emph{Baseline Priority} is only a function of the number of ACP flows and is not impacted by the number of TCP flows. This throughput is $0.8$ Mbps for when we have $2$ ACP flows and goes up to about $5$ Mbps for $20$ ACP flows.

\begin{figure}[tb]             
\begin{center}
\subfloat[\small  ]{\includegraphics[width=.46\linewidth]{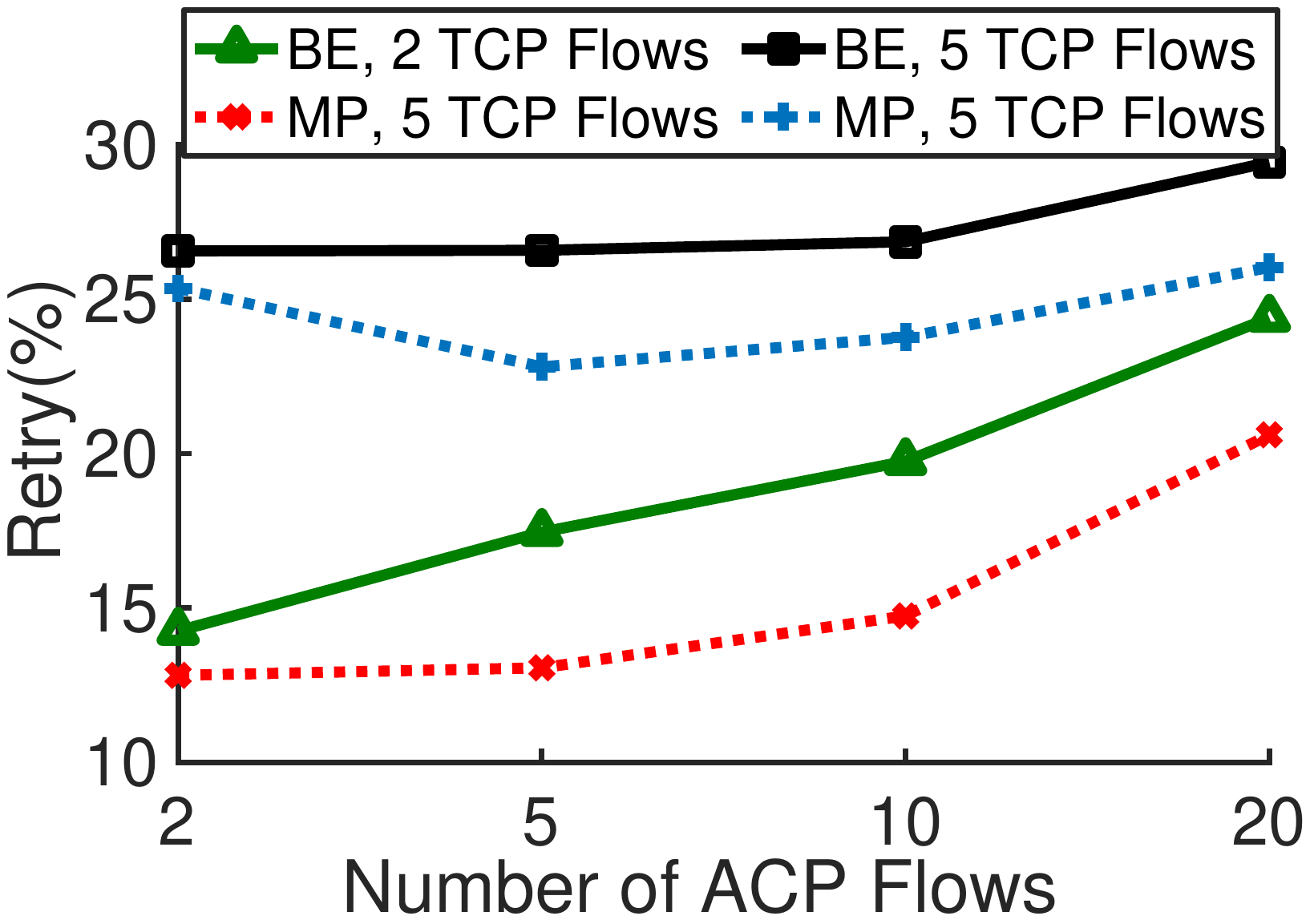}\label{fig:ACPRetryPercent}}
\hspace{0.2in}
\subfloat[\small  ]{\includegraphics[width=.46\linewidth]{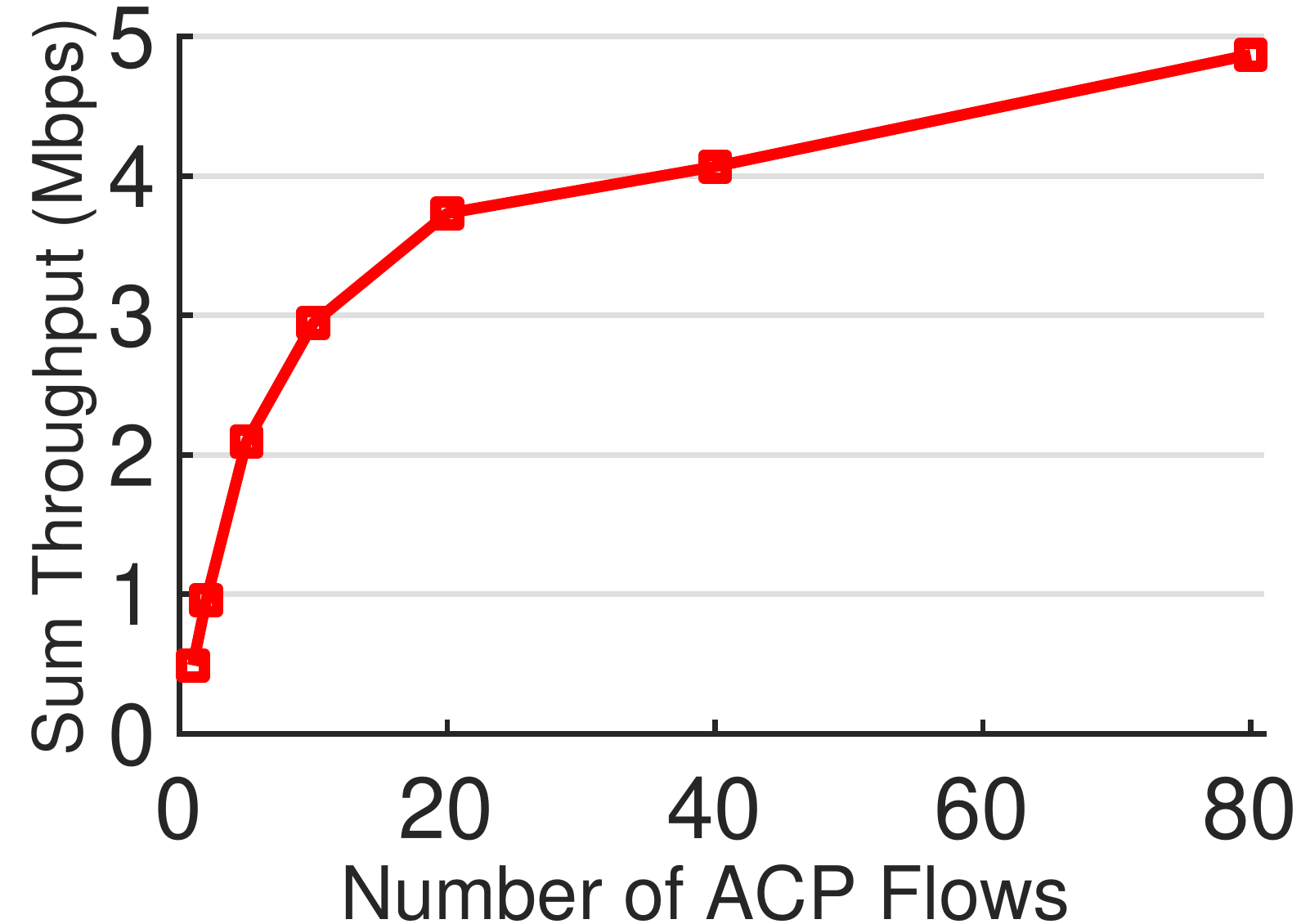}\label{fig:ACPThroughput6M_NoTCP}}
\caption{\small (a) Retry percentages of update packets sent in ACP flows as a function of number of ACP sources. Percentages are shown for \emph{Best Effort} and \emph{Multiaccess Priority}, and for $2$ and $5$ TCP flows. (b) ACP sum throughput in the absence of TCP as a function of the number of ACP flows sharing a $6$ Mbps WiFi link.}
\label{fig:retry_test}
\end{center}
\end{figure}

Further note, for a given number of ACP flows, the sum throughput of TCP flows stays about the same for the configurations BP, MP, and BE. 
For BE, it is within $\approx 2$ Mbps of sum TCP throughputs for \emph{Baseline Priority}. Specifically, for larger numbers of ACP flows, the TCP sum throughput is greater by at most $\approx 1$ Mbps when using \emph{Multiaccess Priority} compared to using BP. When using BE, it is at most $\approx 2$ Mbps higher. TCP throughput benefits in BE because ACP flows have the same access priority as TCP flows. 
The above observation tells us that the significant increases in mean age seen in Figure~\ref{fig:ACPAge} with an increase in the number of TCP flows for a given number of ACP flows, for MP and BE, may not be entirely attributed to TCP's throughput share. 

While an increase in TCP flows doesn't impact the TCP sum throughput, it results in increased MAC layer retries of packets of ACP flows. It also results in ACP flows experiencing large RTTs. Figure~\ref{fig:ACPRetryPercent} shows the packet retry percentages as a function of the number of ACP flows for two and five TCP flows and the configurations of \emph{Multiaccess Priority} and \emph{Best Effort}. Retry percentages increase by about $5\%$ - $10\%$ when the numbers of TCP flows increase from $2$ to $5$. We also see higher retry percentages when there are larger numbers of ACP flows. Also, observe that for a given number of TCP flows, \emph{Best Effort} sees higher retry percentages than MP. This is because having priority has ACP flows see a little less contention over the WiFi multiaccess.

\begin{figure}[tb] 
\vspace*{-0.7em}
\begin{center}
\subfloat[\small Ten ACP flows]{\includegraphics[width=.5\linewidth]{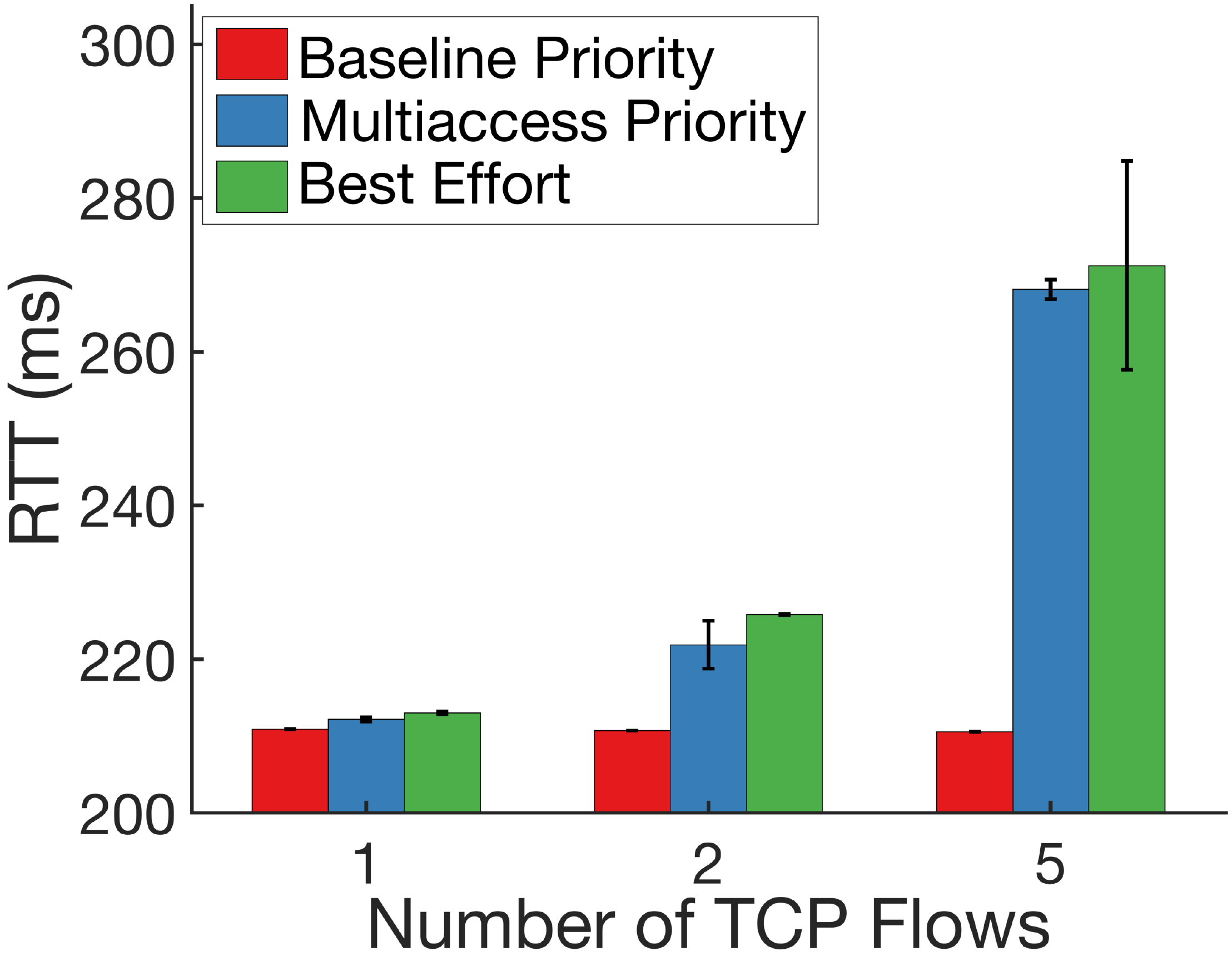}\label{fig:ACPRTTTenFlows}}
\subfloat[\small Twenty ACP flows]{\includegraphics[width=.5\linewidth]{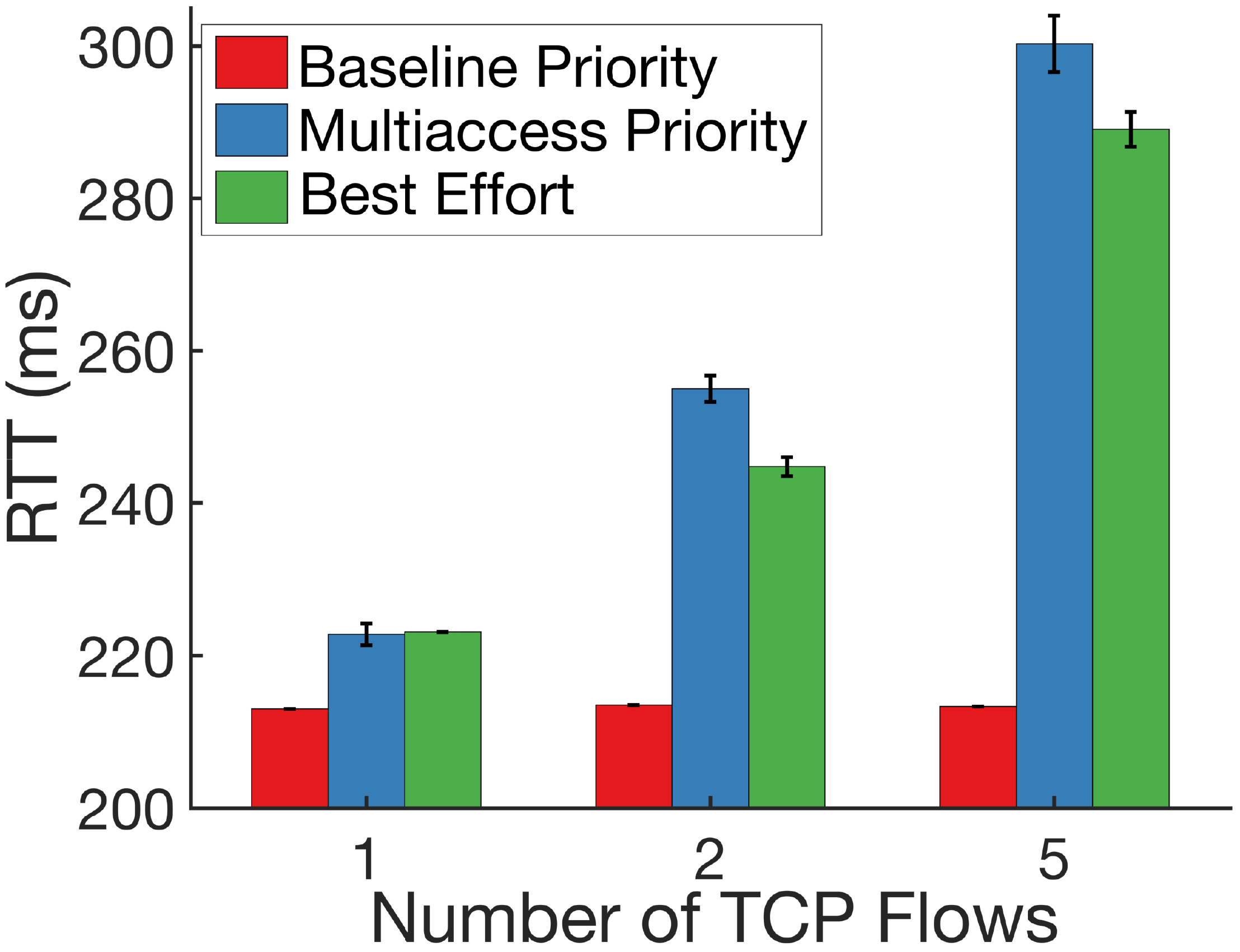}\label{fig:ACPRTTTwentyFlows}}
\caption{\small Mean RTT of ACP flows. For ten and twenty ACP flows, RTT is shown for  \emph{Baseline Priority}, \emph{Multiaccess Priority} and \emph{Best Effort}, and for $1$, $2$, and $5$ coexisting TCP flows.}
\label{fig:ACPRTT}
\end{center}
\end{figure}

We also look at the mean RTTs of updates packets to see the impact of increased MAC layer retries. Figures~\ref{fig:ACPRTTTenFlows} and~\ref{fig:ACPRTTTwentyFlows} show, respectively for $10$ and $20$ ACP flows, significant increases in RTT as the number of TCP flows increase from $1$ to $5$, for MP and BE. These, together with the retry rates, explain the large mean ages observed when using multiaccess in comparison to when using \emph{Baseline Priority}.

\begin{tcolorbox}[title=Summary, breakable, skin=enhanced jigsaw]
\emph{Gains from prioritizing ACP flows vanish quickly with an increase in contention over the shared WiFi multiaccess. The increased contention leads to higher retries and higher RTTs, resulting in higher time-average age.}
\end{tcolorbox}



\smallskip
\noindent \emph{How does the performance of ACP flows sharing a $6$ Mbps WiFi link without interference from TCP flows compare to all flows sharing a $12$ Mbps WiFi link?}
%

\begin{figure}[t]             
\begin{center}
\includegraphics[width = 0.8\linewidth]{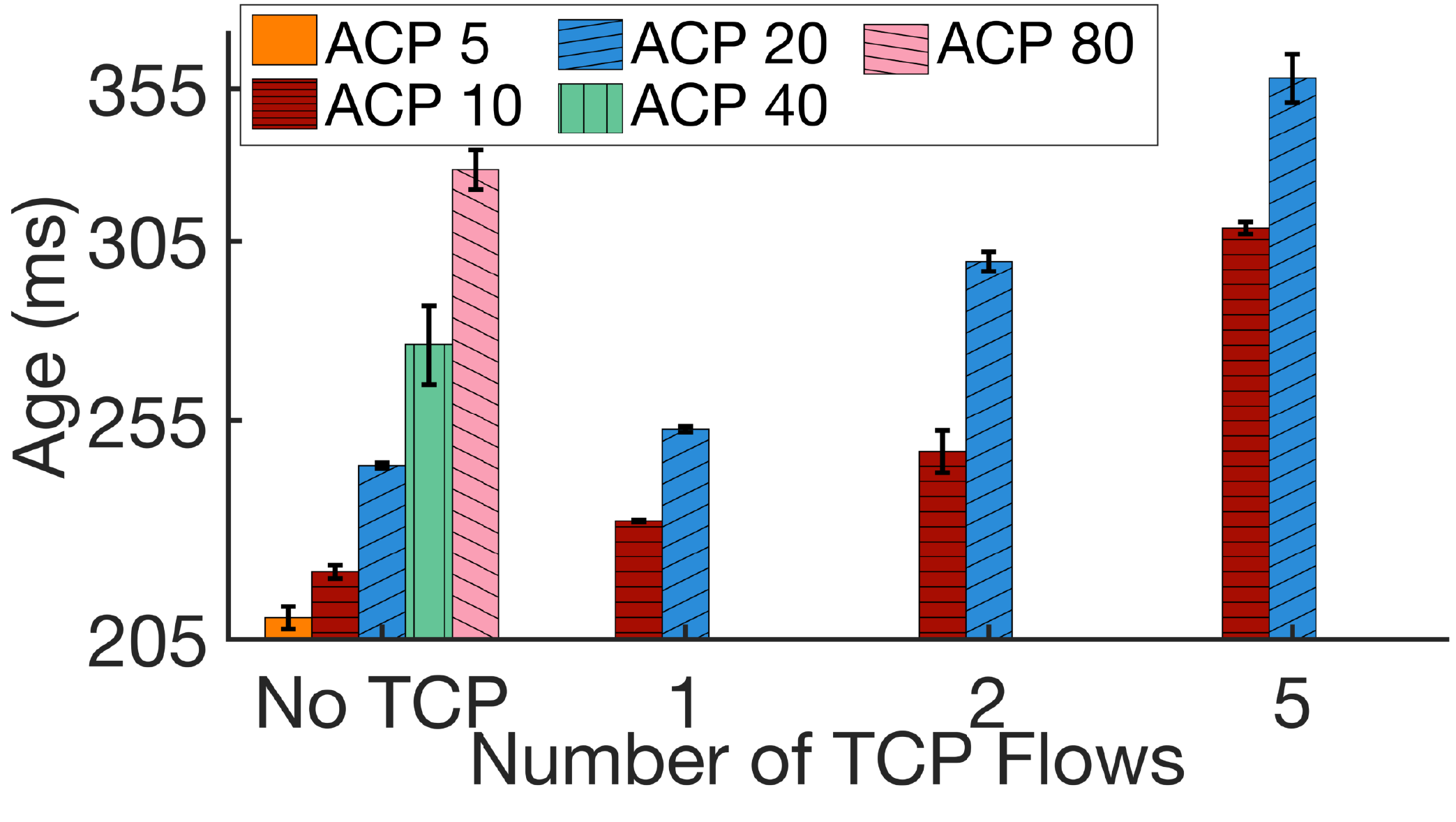}
\caption{\small Mean time-average age for ACP flows with increasing TCP flows. In \emph{No TCP}, all ACP flows share a $6$ Mbps WiFi link. For all other settings, ACP and TCP flows share a $12$ Mbps link in \emph{Multiaccess Priority} configuration.  
}
\label{fig:ACPNoTCPvsTCP}
\end{center}
\end{figure}

Figure~\ref{fig:ACPNoTCPvsTCP} shows the mean age achieved by ACP flows when they share the WiFi multiaccess of rate $6$ Mbps in the absence of TCP flows (labeled \emph{No TCP}) and when the ACP and TCP flows share a $12$ Mbps in MP configuration. 
For the \emph{No TCP} setting, we see that the mean age is $209.43$, $219.18$, $243.79$, $275.15$, $327.41$ ms, respectively for $5$, $10$, $20$, $40$, and $80$ ACP flows. The corresponding sum ACP throughputs (see Figure~\ref{fig:ACPThroughput6M_NoTCP}) are $2$, $2.9$, $3.7$, $4$ and $5$ Mbps. So with $80$ ACP flows sharing the WiFi multiaccess with a link rate of $6$ Mbps, and utilizing almost all of it (a sum throughput of $5$ Mbps), the mean age is $327.41$ ms. Compare these mean ages for the \emph{No TCP} setting with $10$ and $20$ ACP flows under \emph{Multiaccess Priority} when a $12$ Mbps WiFi link is shared with $1$ - $5$ TCP flows (see: Figure~\ref{fig:ACPNoTCPvsTCP}). For $10$ ACP nodes it is $230.7$, $247.25$, and $309$ ms. For $20$ ACP it is $252.76$, $298.9$, and $358.77$ ms, respectively. Clearly, ACP achieves lower age even with $80$ flows and lower link rate (of $6$ Mbps) when compared with $20$ ACP flows coexisting with TCP flows even at a higher link rate of $12$ Mbps. 

For TCP, throughput is the utility of interest. 
For TCP flows sharing a $6$ Mbps WiFi link (without any ACP flows), the sum TCP throughput is $\approx 5.5$ Mbps, which is the expected payload rate after accounting for overheads like packet headers. 
For ACP and TCP flows sharing a $12$ Mbps link, the sum throughput is 5 Mbps ($1$, $2$ or $5$ TCP flows) for when we have $20$ ACP flows and is in the range of $6$ - $6.5$ Mbps for $10$ ACP flows (see Figure~\ref{fig:SumThr}). 
In fact, it is only when we have very few ACP flows, that the TCP sum throughputs are much larger than $5.5$ Mbps. For when TCP shares with only $1$ ACP flow, the sum TCP throughput is as high as $8$ Mbps. With $5$ ACP flows, the sum throughput ranges from $7$ - $7.5$ Mbps.

Additionally, the retry percentages for \emph{No TCP} are $2\%$ for $5$ ACP flows and increase to $17\%$ for $80$ ACP flows (plot not included). Contrast these with the much higher retry rates in \Cref{fig:ACPRetryPercent} for when ACP flows share a $12$ Mbps link with TCP flows. 
We also look at the impact of retry rate on RTTs and find that even the RTTs are smaller, with $20$ ACP flows seeing an RTT less than $220$ ms. 

\begin{tcolorbox}[title=Summary, breakable, skin=enhanced jigsaw]
\emph{When $20$ ACP flows share a $12$ Mbps access with TCP flows, having TCP and ACP flows use non-interfering $6$ Mbps WiFi links is beneficial to both. ACP mean ages are much smaller and TCP gets a higher sum throughput of $5.5$ Mbps.}
\end{tcolorbox}

 \section{Conclusion}
\label{sec:conclusions}
We studied the impact of prioritization on the performance of age-sensitive traffic in the presence of competing network traffic. 
We considered an array of experimental configurations in real-world network settings.
Our results indicate that ACP flows gain from prioritization only when contention over the wireless access from competing traffic is low. The gains are non-existent as the contention increases. 
We also find that a large number of ACP and TCP flows using non-interfering $6$ Mbps WiFi links results in both better throughput and age performance, respectively for TCP and ACP flows, than when the flows share a $12$ Mbps access.

\begin{spacing}{0.92}
\bibliographystyle{IEEEtran} 
\bibliography{IEEEtran,AOI-test}
\end{spacing}
\end{document}